\documentstyle[prd,preprint,tighten,aps,amssymb,newlfont]{revtex}
\begin{document}
\draft
\preprint{
\begin{tabular}{r}
KIAS-P97007\\
JHU-TIPAC 97019\\
DFTT 71/97\\
hep-ph/9711363
\end{tabular}
}
\title{Coherence of neutrino oscillations in the wave packet approach}
\author{C. Giunti}
\address{INFN, Sezione di Torino, and Dipartimento di Fisica Teorica,
Universit\`a di Torino,\\
Via P. Giuria 1, I--10125 Torino, Italy}
\author{C. W. Kim}
\address{Department of Physics $\&$ Astronomy,
The Johns Hopkins University,\\
Baltimore, MD 21218, USA, and\\
School of Physics,
Korea Institute for Advanced Study,
Seoul 130-012, Korea}
\maketitle
\begin{abstract}
The temporal and spatial coherence widths
of the microscopic process by which
a neutrino is detected
are incorporated in the
quantum mechanical wave packet treatment of neutrino oscillations,
confirming the observation
of Kiers, Nussinov and Weiss
that an accurate measurement of the energies
of the particles participating in the detection process
can increase the coherence length.
However,
the wave packet treatment presented here
shows that
the coherence length has an upper bound,
determined by the neutrino energy and
the mass-squared difference,
beyond which the coherence of the oscillation process is lost.
\end{abstract}


A complete understanding of neutrino oscillations
must take into account the localization
of microscopic processes
by which a neutrino is produced and detected.
This localization
is appropriately described by a wave packet
treatment of neutrino oscillations
\cite{Nussinov,BP78,Kayser,GKL91,GKLL93,KNW96,Campagne,GKL97A,KW97}
(different treatments are discussed in \cite{Rich,GS96}).
As the authors of \cite{KW97} noticed,
in the quantum mechanical wave packet approach
presented in \cite{GKL91}
the dependence of the oscillation 
probability on the temporal and spatial coherence widths
of the detection process was neglected.
In this brief report,
we wish to incorporate,
in a simple and straightforward way, 
the temporal and spatial coherence widths
of the detection process
in the quantum mechanical wave packet description
of neutrino oscillations and show that,
as an immediate consequence of this, 
performing an accurate measurement
of the energies of the particles involved
in the detection process leads to an increase of
the coherence length for neutrino oscillations,
as was noticed for the first time in \cite{KNW96}.

Let us consider a neutrino of flavor $\alpha$
produced by a weak interaction process at the
origin of the space-time coordinates
and detected at a distance $L$
after a time $T$\footnote{Taking into account
the spatial and temporal coherence widths of the production and detection
processes,
$x=0$, $t=0$
and
$x=L$, $t=T$
are their average
space-time coordinates.}
by a weak interaction process
capable of detecting  a neutrino of flavor
$\beta$.
As in \cite{GKL91},
we describe the neutrino
propagating between
the production and detection processes
with the
time-dependent state
in the Schr\"odinger picture
\begin{equation}
| \nu_{\alpha}(t) \rangle
=
\sum_{a}
U_{{\alpha}a}^*
\int \mathrm{d}p
\,
\psi_a(p;p_a,\sigma_{pP})
\,
e^{-iE_a(p)t}
\,
| \nu_a(p) \rangle
\,,
\label{01}
\end{equation}
where
$ E_a(p) = \sqrt{ p^2 + m_a^2 } $.
This state is a superposition of
mass eigenstate wave packet states
(labeled by the index $a$)
weighted by the complex-conjugated
elements of the mixing matrix $U$
of the neutrino fields.
For simplicity,
we consider only one space dimension
in the source--detector direction
and
we assume that the mass eigenstate wave functions in momentum space
$\psi_a(p;p_a,\sigma_{pP})$
have the following gaussian form,
which enables us to carry out an analytical calculation
of the transition probability:
\begin{equation}
\psi_a(p;p_a,\sigma_{pP})
=
\left( 2 \pi \sigma_{pP}^2 \right)^{-1/4}
\exp\left[
-
\frac
{ ( p - p_a )^2 }
{ 4 \sigma_{pP}^2 }
\right]
\,,
\label{02}
\end{equation}
where
$p_a$
are the average momenta of the different mass eigenstates
and $\sigma_{pP}$ is the momentum width
of the wave packets.
The average momenta $p_a$ are determined
by the kinematics of the production process
taking into account the masses $m_a$
of the mass eigenstates.
We assume that all the mass eigenstates
are extremely relativistic,
i.e.
$ p_a \gg m_a $.
In this case
$ \Delta E \simeq \Delta p $
and
$\sigma_{pP}$
is determined by the \emph{minimum}
between the energy and momentum
uncertainties of the production process
(a possible dependence of the momentum widths from the index $a$
can be estimated to be very small
and negligible for relativistic neutrinos).
Hence,
from the uncertainty principle
it is clear that $\sigma_{pP}$
is determined by the \emph{maximum}
between the spatial and temporal coherence
widths of the production process.

We assume that the gaussian wave functions (\ref{02})
are sharply peaked around the corresponding average momentum,
i.e.
$
\sigma_{pP}
\ll
E_a^2 / m_a
$,
with
$ E_a \equiv E_a(p_a) $.
Under this condition the energy
$ E_a(p) $
can be approximated by
$
E_a(p)
\simeq
E_a
+
v_a ( p - p_a )
$,
where
$ v_a = p_a / E_a $
is the group velocity of each wave packet.
In this case,
the neutrino wave function in coordinate space
$
| \nu_{\alpha}(x,t) \rangle
=
\langle x | \nu_{\alpha}(t) \rangle
$
is easily calculated after a gaussian integration
to be given by
\begin{equation}
| \nu_{\alpha}(x,t) \rangle
=
\left( 2 \pi \sigma_{xP}^2 \right)^{-1/4}
\sum_{a}
U_{{\alpha}a}^*
\exp\left[
- i E_a t + i p_a x
-
\frac
{ ( x - v_a t )^2 }
{ 4 \sigma_{xP}^2 }
\right]
| \nu_a \rangle
\,,
\label{03}
\end{equation}
with the orthonormal states
$ | \nu_a \rangle $
belonging to the
mass Fock space
($ \langle \nu_a | \nu_b \rangle = \delta_{ab} $).
The width
$\sigma_{xP}$
of the mass eigenstate wave packets in coordinate space
is related to 
the momentum width $\sigma_{pP}$
by the uncertainty relation
$ \sigma_{xP} \, \sigma_{pP} = 1/2 $.
From the considerations presented after Eq.(\ref{02})
it is clear that the value of
$\sigma_{xP}$
is given by the
maximum between the spatial and temporal coherence widths
of the production process.

Let us consider now the detection process
which takes place at a distance $L$ and after a time $T$
from the origin of the space-time coordinates
and is capable of detecting a neutrino
with flavor $\beta$.
In \cite{GKL91}
the amplitude of the flavor changing process
was obtained by projecting the
state
$ | \nu_{\alpha}(L,T) \rangle $,
which describes the propagating neutrino,
on the flavor state
$ | \nu_{\beta} \rangle = \sum_a U_{{\beta}a}^* | \nu_a \rangle $.
This procedure neglects the temporal and spatial coherence widths of
the detection process.
In order to take them into account,
the detected neutrino must be described by
a wave packet state analogous to that in Eq.(\ref{01}):
\begin{equation}
| \nu_{\beta} \rangle
=
\sum_{a}
U_{{\beta}a}^*
\int \mathrm{d}p
\,
\psi_a(p;p_a,\sigma_{pD})
\,
| \nu_a(p) \rangle
\,.
\label{04}
\end{equation}
The value of the momentum width $\sigma_{pD}$
is given, for relativistic neutrinos,
by the minimum between the
energy and momentum uncertainties of the detection process.
The state
$ | \nu_{\beta} \rangle $
does not evolve in time
because it does not represent a propagating neutrino.
The average momenta $p_a$
in the mass eigenstate gaussian wave functions
describing the detected neutrino
are the same as those of the corresponding
mass eigenstate gaussian wave functions
describing the neutrino propagating between
production and detection.
This property is a consequence of causality:
after the average momenta $p_a$
have been determined by the kinematics of
the production process,
the mass eigenstates propagate
between the two processes and
determine the kinematics in the detection process.
For example,
the moduli of the average momenta $p_a$
of the mass eigenstate wave packets of
a muon neutrino produced in pion decay at rest
are fixed by the kinematics of the process.
When this neutrino is detected,
for example,
by scattering with a nucleus at rest,
each mass eigenstate
determines with its
average momentum $p_a$ and its mass $m_a$
a different value for the
momenta of the recoil particles.
Neutrino oscillations are observed if
the different mass eigenstates
are detected coherently,
i.e.
if the differences of the energies and momenta
of the recoil particles corresponding to
different mass eigenstates
are smaller than the energy and momentum uncertainty
of the detection process.

Taking into account that the detection process
takes place at a distance $L$
from the origin of the coordinates,
the wave function in coordinate space describing
the detected neutrino is given by
\begin{equation}
| \nu_{\beta}(x-L) \rangle
=
\left( 2 \pi \sigma_{xD}^2 \right)^{-1/4}
\sum_{a}
U_{{\beta}a}^*
\exp\left[
i p_a (x-L)
-
\frac
{ ( x - L )^2 }
{ 4 \sigma_{xD}^2 }
\right]
| \nu_a \rangle
\,,
\label{05}
\end{equation}
where
$\sigma_{xD}$
is defined by the uncertainty relation
$ \sigma_{xD} \, \sigma_{pD} = 1/2 $.
Hence,
the value of $\sigma_{xD}$
is given by the maximum between the spatial and temporal
coherence widths of the detection process.

The amplitude of the flavor changing process
is given by the overlap
\begin{equation}
A_{\alpha\beta}(L,T)
=
\int \mathrm{d}x
\,
\langle \nu_{\beta}(x-L) | \nu_{\alpha}(x,T) \rangle
\,.
\label{06}
\end{equation}
The integral over $x$ is gaussian and leads to
\begin{equation}
A_{\alpha\beta}(L,T)
=
\sqrt{ \frac{2\sigma_{xP}\sigma_{xD}}{\sigma_{x}^2} }
\sum_{a}
U_{{\alpha}a}^*
U_{{\beta}a}
\exp\left[
- i E_a T + i p_a L
-
\frac
{ ( L - v_a T )^2 }
{ 4 \sigma_{x}^2 }
\right]
\,,
\label{07}
\end{equation}
with
\begin{equation}
\sigma_{x}^2
\equiv
\sigma_{xP}^2
+
\sigma_{xD}^2
\,.
\label{08}
\end{equation}
This relation is very important in that
it clearly shows that the width
$ \sigma_{x} $
which determines the coherence of the flavor changing process
depends on the spatial and temporal coherence widths
of both the production and detection processes.
The amplitude
(\ref{07})
has the same form as that in \cite{GKL91},
with an important difference: the width
$ \sigma_{x} $
in \cite{GKL91}
was determined only by the production process,
whereas in Eq.(\ref{07}) also the
detection process is properly taken into account.

In order to calculate the oscillation probability,
the mass eigenstate energies can be approximated by
\begin{equation}
E_{a} \simeq  E + \xi \, \frac{m_{a}^2}{2E}
\,,
\label{09}
\end{equation}
where
$E$
is the energy determined by the kinematics of the production process
for a massless neutrino
and
$\xi$
is a dimensionless quantity that is determined
by energy-momentum conservation
in the production process
to first order in $m_{a}^2/E^2$.
The quantity $\xi$ is typically of order unity;
for example, in neutrino production
by pion decay at rest we have $\xi\simeq0.2$.
With this approximation we have
$
p_{a}
\simeq
E
-
\left( 1 - \xi \right)
m_{a}^2/2E
$
and
$
v_{a}
\simeq
1
-
m_{a}^2/2E^2
$.
From Eqs.(\ref{07}) and (\ref{09}), 
the probability of
$\nu_\alpha\to\nu_\beta$
transitions is given by
\begin{eqnarray}
&&
P_{\alpha\beta}(L,T)
\propto
\sum_{a,b}
U_{{\alpha}a}^{*} \,
U_{{\beta}a} \,
U_{{\alpha}b} \,
U_{{\beta}b}^{*} \,
\exp
\left\{
- i
\frac{ \Delta{m}_{ab}^2 }{ 2 E }
\,
[ \xi T + (1-\xi) L ]
\right\}
\nonumber
\\
&&
\hspace{5.5cm}
\times
\exp
\left\{
-
\frac{
( L - v_a T )^2
+
( L - v_b T )^2
}
{ 4 \sigma_{x}^2 }
\right\}
\,,
\label{10}
\end{eqnarray}
with
$
\Delta{m}_{ab}^2
\equiv
m_a^2 - m_b^2
$.

In principle,
$P_{\alpha\beta}(L,T)$
is a measurable quantity,
but  in all realistic experiments
the distance $L$ is a fixed and known quantity,
whereas the time $T$ is not measured
and can have any value,
because the source and the detector typically operate for
times much longer than the oscillation times
$ 4 \pi E / \Delta{m}^2_{ab} $.
Therefore,
the quantity that is measured in all experiments is
the oscillation probability
$P_{\alpha\beta}(L)$
at a fixed distance $L$,
given by the time average of
$P_{\alpha\beta}(L,T)$.
After integrating over $T$
and imposing the normalization condition
$
\sum_{\beta}
P_{\alpha\beta}(L)
=
1
$,
we obtain
\begin{equation}
P_{\alpha\beta}(L)
=
\sum_{a,b}
U_{{\alpha}a}^{*}
U_{{\beta}a}
U_{{\alpha}b}
U_{{\beta}b}^{*}
\exp
\left[
- 2 \pi i
\,
\frac{ x }{ L_{ab}^{\mathrm{osc}} }
-
\left(
\frac{ x }{ L_{ab}^{\mathrm{coh}} }
\right)^2
\right]
F_{ab}
\,,
\label{11}
\end{equation}
where
\begin{equation}
L_{ab}^{\mathrm{osc}}
=
\frac
{ 4 \pi E }
{ \Delta{m}_{ab}^2 }
\,,
\qquad
L_{ab}^{\mathrm{coh}}
=
\frac
{ 4 \sqrt{2} \sigma_{x} E^2 }
{ \left| \Delta{m}_{ab}^2 \right| }
\label{12}
\end{equation}
are the oscillation wavelengths and coherence lengths,
respectively,
and
\begin{equation}
F_{ab}
=
\exp
\left[
-
2 \pi^2
( 1 - \xi )^2
\left(
\frac{ \sigma_{x} }{ L_{ab}^{\mathrm{osc}} }
\right)^2
\right]
\,.
\label{13}
\end{equation}
Equation (\ref{11}) contains,
in addition to
the usual expression for the
neutrino oscillation probability
(the first term in the exponential with $F_{ab} =1$),
two additional factors, the first being 
the second term in the exponential
that takes into account the coherence
of the contributions of different
mass eigenstate wave packets
and the second being  the additional factor $F_{ab}$.
The factor $F_{ab}$ is 
equal to unity if
$
\sigma_{x}
\ll
| L_{ab}^{\mathrm{osc}} |
$,
which is a necessary condition
for the observation of neutrino oscillations
that must be satisfied by any  realistic experiment.
If this condition is not satisfied,
the interference among different
mass eigenstate wave packets is washed out
and only a constant
transition probability
$
P_{\alpha\beta}
=
\sum_{a}
| U_{{\alpha}a} |^2
| U_{{\beta}a} |^2
$
can be observed.

It is important to notice that
the integration over the oscillation time
is a crucial step
in the wave packet treatment of
neutrino oscillations in space,
because it allows us  to eliminate the time degree
of freedom,
which is not measured in realistic experiments.
This elimination is done at
the classical level
by integrating the probability,
which is a classical quantity,
and without
any unphysical assumption on the
energies and momenta of the mass eigenstates,
which are determined by the production process
\cite{Winter81}.
An unphysical assumption is,
for example,
the imposition of equal energy to
the different mass eigenstates,
i.e. $\xi=0$,
which would eliminate
the time dependence of the oscillatory term
in the probability (\ref{10}).

From Eqs.(\ref{08}) and (\ref{12}) one can see that
the coherence length
$L_{ab}^{\mathrm{coh}}$
is proportional to
$\sigma_x\equiv\sqrt{\sigma_{xP}^2+\sigma_{xD}^2}$.
Hence,
$L_{ab}^{\mathrm{coh}}$
is dominated by the largest among the
temporal and spatial coherence widths
of the production and detection process.
In particular,
a precise determination
of the energies
(or momenta)
of the particles involved
in the detection process
implies a small $\sigma_{pD}$
and a large $\sigma_{xD}$,
leading to a large coherence length
$
L_{ab}^{\mathrm{coh}}
\simeq
4 \sqrt{2} \sigma_{xD} E^2
/
\left| \Delta{m}_{ab}^2 \right|
$
(if $\sigma_{pD}\ll\sigma_{pP}$).
Hence,
as noted in \cite{KNW96},
the coherence length
can be increased by measuring accurately
the energies
of the particles participating in
the detection process.
However,
even if,
at least in principle,
the coherence length can be increased without limit
by an extremely precise measurement of the energies
of the detection (or production) particles,
the oscillations lose the coherence for
$
\sigma_{x}
\gtrsim
L_{ab}^{\mathrm{osc}}
$,
because the factor
$F_{ab}$
becomes important and has the effect of
suppressing the interference of
$\nu_a$ and $\nu_b$
for
$
\sigma_{x}
\gg
L_{ab}^{\mathrm{osc}}
$.
Physically this is due to the fact
that the spatial or temporal width
of the detection (or production)
process becomes larger than the oscillation length,
leading to the washout of the oscillations\footnote{Notice
that a similar effect is obtained
if the distance $L$ from the source is not known
and the time-dependent probability
$P_{\alpha\beta}(T)$
is obtained by averaging
$P_{\alpha\beta}(L,T)$
over $L$.
Such a situation is realized,
for example,
in the calculation of the effects of neutrino oscillations
in the early universe.}.
In order to understand
the reason of this washing-out
let us consider, for example,
the case in which the detection process has a small spatial coherence width and a large
temporal coherence width that dominates the total coherence width $\sigma_x$.
As shown by Eq.(\ref{10}), in this case the interference between
the mass eigenstates $a$ and $b$
is not suppressed
if the average detection
time $T$
does not differ from $L/v_a$ and $L/v_b$ by more than $\sim\sigma_x$.
This is due to the fact that the modes of the detection process
corresponding to the different mass eigenstates oscillate 
coherently during a time interval of width $\sigma_x$ around $T$,
even when the mass eigenstate wave packets are far from
$L$\footnote{Notice that if $\sigma_{xD}$ is dominated by the temporal
coherence width of the detection process, the gaussian approximation in
(\ref{05})
for the wave function of the detected neutrino introduces an unphysical
interaction for times smaller than the time of arrival of the propagating
neutrino to the detector.
In this case a more realistic approximation can be
obtained
by inserting in (\ref{05}) and (\ref{06})
a $\theta(x-L+\tilde\sigma_{xD})$,
where
$\tilde\sigma_{xD}$ is the spatial coherence width of the detection process.
This approximation requires
a numerical solution of the integral in (\ref{06}) and
will be discussed in detail elsewhere \cite{GKL98}.}.
Then it is clear that if $\sigma_x$ is larger that the oscillation length,
the phase of the interference term depends crucially on the value of $T$.
In this case,
if $T$ is not measured the average
of the transition probability over $T$ washes out the interference.

Therefore,
using Eq.(\ref{12})
one can see that
the coherence length
has the upper bound
\begin{equation}
L_{ab}^{\mathrm{coh}}
\lesssim
E \, (L_{ab}^{\mathrm{osc}})^2
=
\frac
{ 16 \pi^2 E^3 }
{ \left( \Delta{m}_{ab}^2 \right)^2 }
\label{14}
\end{equation}
beyond which the coherence length loses its meaning,
because the coherence of the process is lost
for any value of the distances $L$.

In conclusion,
we have shown that the temporal and spatial coherence widths
of the detection process
can easily be incorporated in the
quantum mechanical wave packet treatment of neutrino oscillations.
As a result,
we confirm the observation presented in \cite{KNW96}
that an accurate measurement of the energies
(or momenta)
of the particles participating in the detection process
can increase the coherence length.
However,
the wave packet treatment presented here
shows that
the coherence length cannot be increased beyond the
upper bound given by Eq.(\ref{14})
without losing the coherence of the oscillation process.


\end{document}